\documentclass[10pt,letterpaper]{article}
\usepackage{opex3}
\usepackage{ae}
\usepackage{amsmath,bm}
\usepackage{graphicx}

\newcommand{\avg}[1]{\langle #1 \rangle}
\newcommand{\swap}{\gamma_\text{sw}}
\begin{document}

\title{Single mode quadrature entangled light from room temperature atomic vapour}
\bibliographystyle{osajnl}

\author{W. Wasilewski$^{1,*}$, T. Fernholz$^1$, K. Jensen$^1$, L. S. Madsen$^1$, H. Krauter$^1$, C. Muschik$^2$ and E. S. Polzik$^1$}
\address{$^1$ Niels Bohr Institute, Danish Research Foundation Center for Quantum Optics (QUANTOP), Blegdamsvej 17, DK-2100 Copenhagen, Denmark}
\address{$^2$ Max-Planck--Institut f\"ur Quantenoptik, Hans-Kopfermann-Strasse, D-85748 Garching, Germany}
\email{$^*$ wwasil@nbi.dk}

\begin{abstract}
 We analyse a novel squeezing and entangling mechanism which is due to correlated Stokes and anti-Stokes photon forward scattering in a multi-level atom vapour. Following the proposal we present an experimental demonstration of 3.5 dB pulsed frequency nondegenerate squeezed (quadrature entangled) state of light using room temperature caesium vapour. The source is very robust and requires only a few milliwatts of  laser power. The squeezed state is generated in the same spatial mode as the local oscillator and in a single temporal mode. The two entangled modes are separated by twice the Zeeman frequency of the vapour which can be widely tuned.
The narrow-band squeezed light generated near an atomic resonance can be directly used for atom-based quantum information protocols. Its single temporal mode characteristics make it a promising resource for quantum information processing.
\end{abstract}


\section{Introduction}

Quadrature entangled states of light occupy an important place in modern quantum information processing, with examples inluding quantum cryptography \cite{RalphPRA00}, teleportation \cite{Furusawa98}, computation \cite{LloydPRL99} and last but not least, improving Quantum Non-Demolition (QND) measurements \cite{KuzmichPRL00}. Although continuos-variable entanglement and squeezing is a fragile resource easily washed out by optical losses, proposals for continuous-variable error correction \cite{BraunsteinNat98,LloydPRL98}, entanglement purification \cite{DuanPRL00} and distillation \cite{BrowEisePRA03} have been developed.

A sub-threshold optical parametric amplifier (OPA) first used to generate quadrature entangled light in \cite{Ou92} remains the mainstream approach \cite{SchoriPRA02,VahlbruchPRL08}. This approach offers design flexibility but at the same time requires relatively complicated and sensitive setup consisting of two or more stabilized cavities. 
The other possibilities include the usage of four wave mixing (4WM) \cite{BoyerSci08}. This and similar methods however generate multimode temporal and/or spatial entanglement, that is the output light of the amplifier contains pairs of frequencies which are independently quadrature entangled.
This may have adverse consequences for some applications. For example, if multiple modes are used in quantum cryptography,
modes not detected by the legitimate addressee of the information can carry the same signal. In principle they can be detected unnoticed, compromising the safety of the protocol. Also frequently a multimode character of the squeezed light is a disadvantage for the protocols involving photon counting, such as entanglement purification \cite{BrowEisePRA03} or preparation of squeezed single-photon states \cite{OurjoumtsevSci06,NeergaardNielsenPRL06,SasakiPRA06}.


In this paper we report the proposal and experimental observation of a high purity entangled state which at the same time is a pure two-mode squeezed state. The source is based on an interaction of off-resonant driving light with the room-temperature spin polarised caesium vapour placed in dc magnetic field in relaxation-protected environment. The squeezed field is naturally compatible with atomic memories based on the same alkali atom as the source. The non-classical state of light is generated in a pair of frequency sidebands shifted by the Zeeman frequency around the driving field (the carrier) with a unique temporal envelope and the same spatial mode as the driving light beam.
The frequency of the entangled modes can be tuned by magnetic field and the bandwidth of squeezing can be adjusted by changing the driving light power or the optical depth of the atomic sample. In the present experiment an ultra-narrow band squeezing with approximately $1$ kHz bandwidth has been generated. These features make it an attractive alternative to existing sources for the applications in quantum information processing (QIP), especially those vulnerable to the spurious modes.

Our method requires almost no alignment and three diode lasers, each of a few mW power, two for optical pumping and one for off-resonant driving.

\section{Model of the interaction}

\begin{figure}[t]
\centering
\includegraphics{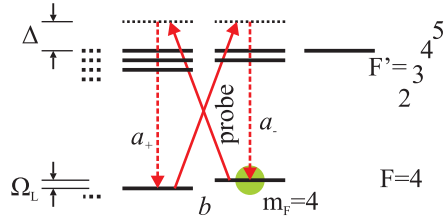}
\caption{Off-resonant double $\Lambda$ interaction in Caesium in the presence of the magnetic field. The interaction is driven by a strong, linearly polarised probe beam detuned by $\Delta= 0.85$GHz from the extreme atomic resonance. The ground state levels are split by $\Omega_L=322$kHz. The collective scattering leads to weak atomic excitation to the $m_F=3$ state described by an annihilation operator $\hat b$ and $\Omega_L$ sidebands of the probe light described by annihilation operators $\hat a_+$ and $\hat a_-$. 
}
\label{fig:levels}
\end{figure}
Let us consider an ensemble of atoms (Fig.~\ref{fig:levels}) with the total ground state angular momentum $F>1/2$ driven by a strong off-resonant linearly-polarized probe light. Prior to the interaction we optically pump all atoms into the extreme magnetic sublevel $m_{F}=F$ in the electronic ground state. The orientation is thus orthogonal to both the propagation and polarization direction of the driving field. The probe light is seen as a superposition of $\sigma_+$ and $\sigma_-$ polarizations in the quantisation basis.

Qualitatively generation of squeezed light by such system can be explained via correlations between forward scattered  $\pi$-polarized photons. The driving light leads to the anti-Stokes scattering into $m_{F}=F-1$ followed by Stokes scattering back to $m_{F}=F$. In the alkali atoms, for the polarization of the driving field perpendicular to the orientation, the second transition is more probable and therefore almost each anti-Stokes photon is accompanied by the Stokes twin with anticorrelated phase. This property is mediated by the collective atomic excitation as in the quantum repeater utilizing DLCZ protocol \cite{DuanNature01}. The correlation persists over a period of several milliseconds, the typical life time of the atomic ground state coherences $T_2$ in our experiment. If the entire process happens on a timescale shorter than $T_2$, then the emitted light is in a single mode. The more photon-atom pairs are created and the more atomic excitations are released into light, the higher is the degree of squeezing which is defined by the number of scattered photons. A general theory of off-resonant interaction of this type neglecting atomic motion has been discussed in \cite{KupriyanovPRA05} and broadband squeezing was predicted in \cite{MishinaNATO06}.

The interaction between light and atoms in our system can be described using two creation operators --- $\hat a^\dagger$ for the scattered photons and $\hat b^\dagger$ for the collective atomic excitation into the magnetic sublevel $m_{F}=F-1 $.  The Hamiltonian of interaction can be written as:
\begin{equation}\label{Eq:HQND+}
    \hat H_\text{int}=-\hbar\chi_a \hat a^\dagger {\hat b}^\dagger+ \hbar\chi_p \hat a^\dagger {\hat b} + \text{H.c.}\,,
\end{equation}
where $\chi_a$ and $\chi_p$ are the coupling constants, describing Stokes and anti-Stokes scattering. At the elementary interaction level $\chi_a \hat a^\dagger \hat b^\dagger+ \text{H.c.}$ describes the active part of the interaction, that is photon-atom entanglement, while $\chi_p \hat a^\dagger \hat b + \text{H.c.}$ describes the passive part of the interaction, that is a beam splitter - like exchange of excitations between photons and atoms.


In case of realistic multilevel atoms
the passive and active coupling $\chi_p$ and $\chi_a$ in Eq.\ \eqref{Eq:HQND+} can have substantially different magnitudes. The interaction Hamiltonian can be written as $\hat H_\text{int}=\hbar\chi (\hat p_a \hat p_b+\xi^{2}\hat x_a \hat x_b)$ where $\chi=\chi_p+\chi_a$, $\xi^2=(\chi_{p}- \chi_{a})/(\chi_{p}+\chi_{a})={14a_2}/{a_1}$, $a_1$, $a_2$ are the vector and tensor parts of the atomic polarizability, and $\hat p_a=(\hat a-\hat a^\dagger)/i\sqrt2$ and $\hat p_b=(\hat b-\hat b^\dagger)/i\sqrt2$ being $p$ quadrature operators associated with light and atoms (for details of the Hamiltonian derivation see the Appendix).
For short interaction times for alkali atoms with $F>1/2$ the second term in $\hat H_\text{int}$ can be neglected and the Hamiltonian reduces to the quantum nondemolition (QND) Hamiltonian $\hat H_\text{QND}=2\hbar\chi \hat p_a \hat p_b$ extensively used for spin squeezing, quantum memory and teleportation protocols \cite{KuzmichPRL00,ShersonRev06}. For atoms with $F=1/2$ the passive and active couplings in Eq.\ \eqref{Eq:HQND+} have equal magnitudes $\chi_a=\chi_p$ and the Hamiltonian is always of the QND type.

However for the stronger coupling/longer interaction time with alkali atoms the complete Hamiltonian Eq.\ \eqref{Eq:HQND+} has to be applied which leads to new attractive dynamics.
If the passive part of the interaction prevails $\chi_p>\chi_a$, which is the case analyzed in this paper, $\xi$ is real and the interaction leads to swapping of the quantum states between light and atoms as discussed below. On the contrary the case of $\chi_a>\chi_p$, i.e. of imaginary $\xi$ leads to entanglement between light and atoms.


 For a rigorous derivation of the atom-photon dynamics we start with the Hamiltonian in the presence of the magnetic field. As illustrated in Fig.~\ref{fig:levels} the ground state sublevels experience Zeeman splitting with the Larmor frequency $\Omega_L$. Therefore the Stokes scattering produces a photon in an upper sideband of the probe, described by the creation operator $\hat a_+^\dagger$, while the anti-Stokes scattering couples to a lower sideband of the probe, described by the creation operator $\hat a_-^\dagger$. Thus the interaction Hamiltonian is reexpressed in the following way:
\begin{equation}\label{Eq:Htot}
\hat H_\text{int}=\frac{\hbar}{\sqrt L}\int_0^L dz \left(-\chi_a \hat a_-^\dagger \acute{\hat b}^\dagger + \chi_p \hat a^\dagger_+ \acute{\hat b}\right) + \text{H.c.}
\end{equation}
 $L$ is the length of the cell and we have omitted the space and time dependence of operators for brevity.
The operator $\acute{\hat b}$ annihilates an atomic excitation in an atomic slice around a given $z$:
\begin{equation}
\acute{\hat b}=\frac{e^{i\Omega_L t}}{n}\sqrt\frac{N_a}{L}\sum_{k=1}^n |m_F=4\rangle_k\langle m_F=3|_k
\end{equation}
where $N_a$ is the total number of atoms in the cell, $n$ is the number of atoms in the slice and $k$ indexes the atoms. In the limit of almost all atoms residing in the $m_F=4$ state $\acute{\hat b}$ is a bosonic operator, $[\acute{\hat b}(z),\acute{\hat b}^\dagger(z)]=\delta(z-z')$.
Analogously, $\hat a_+$ and $\hat a_-$ denote an annihilation operator for the upper or lower sideband photonic mode around a given $z$ at a certain time with $[\hat a_+(z,t),\hat a_+^\dagger(z,t')]=\delta(t-t')$.

 We derive the input-output relations from this Hamiltonian under two assumptions. We assume that the light passes through the cell on the timescale much shorter than the total time of the interaction or the atomic transient time. Next we assume that the atoms see an average over the volume polarization of the light since they move fast compared to the evolution of their internal state. Therefore the collective atomic mode interacting with light is uniform across the cell with the corresponding annihilation operator
$ \hat b=\int_0^L dz\, \acute{\hat b}(z)/\sqrt L. $
We then obtain for the light modes:
\begin{align} \label{Eq:aout=,1cell}
\hat a'_+(t)&=\hat a_+(t) - i\chi_p \hat b(t),  \nonumber\\
\hat a'_-(t)&=\hat a_-(t) + i\chi_a \hat b^\dagger(t),
\end{align}
where $\hat a_\pm(t)=\hat a_\pm(z=0,t)$ and
$\hat a'_\pm(t)=\hat a_\pm(z=L,t)$ denote the operators at the input/output plane of the cell,
and we assume that the passage is instantaneous. Note, that the state of light inside the cell changes linearly with the coordinate $z$ from $\hat a_\pm(t)$ to $\hat a'_\pm(t)$.

To calculate the atomic state evolution we integrate the Heisenberg equation for the local atomic polarisation  $\acute{\hat b}$ over the length of the cell to obtain an equation describing the evolution of the entire ensemble:
\begin{equation} \label{Eq:db=,1cell}
\frac{\partial \hat b(t)}{\partial t}= - i\chi_p \hat a_+(t) + i \chi_a \hat a_-^\dagger(t) -\swap\hat b(t),
\end{equation}
where  the factor $\swap=|\chi_p|^2/2-|\chi_a|^2/2$ takes into account the linear change of $a_+$ and $a_-$ along the cell, as follows from Eq.\ \eqref{Eq:aout=,1cell} . We will see that $\swap$ is the rate at which the initial state of the atoms decays and is replaced by the state of the incoming light, hence we refer to it as a {\it swap rate}.
This equation can be readily solved and yields the single cell input-output relations. However the result is complicated and not useful for our current purpose, because we have disarrayed the simple structure of the Hamiltonian by applying the magnetic field. Fortunately, similarly to the QND case this can be rectified by letting the light interact with two atomic ensembles in series \cite{Julsgaard2001,HammererRMP09}. The ensembles are placed in equal magnetic fields, with the atoms pumped into opposite extreme magnetic states. Then the upper and lower sidebands interchange their roles.

To obtain the combined two-cell input-output relations we write the Eqs.~\eqref{Eq:aout=,1cell} and \eqref{Eq:db=,1cell} for both cells, using the output light of the first cell as the input for the second cell. Instead of sideband operators $a_+$ and $a_-$ we shall use the sine and cosine combinations $\hat a_c=(\hat a_++\hat a_-)/\sqrt2$ and $\hat a_s=i(-\hat a_++\hat a_-)/\sqrt2$. For the atoms we use collective annihilation operators defined as $\hat b_c=(\hat b_1+\hat b_2)/\sqrt2$ and $\hat b_s=i(-\hat b_1+\hat b_2)/\sqrt2$ where $\hat b_1$ and $\hat b_2$ are annihilation operators for cell 1 and 2 respectively. We arrive at:
\begin{align}
\hat a''_c(t)&=\hat a_c(t)-i\chi_p \hat b_c(t) + i\chi_a \hat b_c^\dagger(t), \nonumber\\
\frac{\partial \hat b_c(t)}{\partial t}&=-i \chi_p \hat a_c(t) +i \chi_a \hat a^\dagger_c(t)
-\swap\hat b_c(t),
\end{align}
where $''$ denotes the output face of the second cell. We also get an identical pair of equations for $\hat a_s$ and $\hat b_s$ operators. Below we shall focus on the equations for $c$ operators implicitly using the fact, that the other solutions will be identical. Integrating the above equations gives:
\begin{subequations}\label{Eq:iotime}\begin{align}
\hat a''_c(t)=&\hat a_c(t)-{2\swap}\int_0^t dt' e^{-\swap(t-t')} \hat a_c(t') 
+ie^{-\swap t} \left(-\chi_p \hat b_c(0) + \chi_a \hat b_c^\dagger(0)\right), \label{Eq:iolight}\\
\hat b_c(t)=
&i\int_0^t dt' e^{-\swap(t-t')} \left(-\chi_p \hat a_c(t') +\chi_a \hat a^\dagger_c(t')\right) 
+e^{-\swap t} \hat b_c(0).  \label{Eq:ioatoms}
\end{align}\end{subequations}

Those equations can be cast into a simple form once suitable temporal modes for the light are introduced. For probe pulses of duration $T$ we choose:
\begin{align}
\hat a_c(t)&=\frac{\hat X_L+i\hat P_L}{\sqrt2} N_\text{in} e^{\swap t} + \ldots, &
\hat a''_c(t)&=\frac{\hat X'_L+i\hat P'_L}{\sqrt2} N_\text{out} e^{-\swap t} + \ldots,
\end{align}
where $\hat X_L$ and $\hat P_L$ are quadrature operators for the new modes of light, $N_\text{in}$ and $N_\text{out}$ are normalization factors while the dots stand for contribution from modes orthogonal to $N_\text{in} e^{\swap t}$ and $N_\text{out} e^{-\swap t}$ respectively. We also find it convenient to introduce the quadrature operators $\hat X_A$ and $\hat P_A$ for the atomic state:
\begin{align}
\hat b_c(0)&=\frac{\hat X_A+i \hat P_A}{\sqrt2}, &
\hat b_c(T)&=\frac{\hat X'_A+i \hat P'_A}{\sqrt2}.
\end{align}
With these notations the input-output relations reduce to:
\newcommand{\transm}{\sqrt{1-\xi^{2}\kappa^2}}
\begin{align}\label{Eq:io}
\hat X_A'=&\transm\hat X_A + \kappa \hat P_L, &
\hat X_L'=&\transm\hat X_L + \kappa \hat P_A, \nonumber \\
\hat P_A'=&\transm\hat P_A - \xi^{2}\kappa \hat X_L, &
\hat P_L'=&\transm\hat P_L - \xi^{2}\kappa \hat X_A.
\end{align}
where $\kappa=1/\xi\sqrt{1-\exp(-2\swap T)}$ is the coupling constant. 
\begin{figure}
\centering
\includegraphics{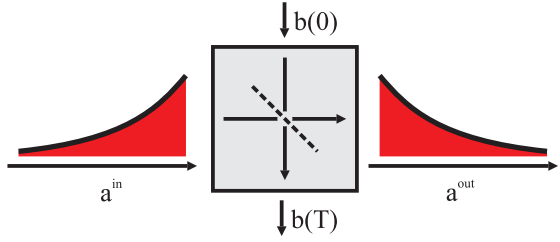}
\caption{Pictorial representation of the interaction. The light in the exponentially rising mode at the input of the cell is mapped onto the atoms, and it assumes a form of an exponentially falling mode onto which the atoms have contributed.}
\label{fig:IO}
\end{figure}
These input-output relations have a simple and interesting form. They describe a beam splitter transformation with transmission $1-\xi^{2}\kappa^2$ and squeezing of the inputs by a factor of $1/\xi$. This is schematically depicted in Fig.~\ref{fig:IO}. In particular after a long enough interaction time $T\gg\swap^{-1}$ the initial state of the atoms is squeezed and mapped onto the light, and vice versa. If the atoms are initially in the coherent spin state (CSS), the output light will be in a squeezed state with a squeezing factor equal $1/\xi$, i.e. Var$(\hat P'_L)=\xi^{2}$. Note that the above equations describe either the cosine combination of the light sidebands and the symmetric excitation of the two ensembles or the sine combination and the antisymmetric excitation. Therefore both the $\hat P'_{L,c}$ and $\hat P'_{L,s}$ will be squeezed after the interaction. This is equivalent to generating quadrature entanglement between upper and lower sideband of the probe pulse since
$2\text{Var}(\hat P'_{L,c})+2\text{Var}(\hat P'_{L,s})=\text{Var}(\hat P_{+}+\hat P_{-})+\text{Var}(\hat X_{+}-\hat X_{-})<2$ where + and - denote upper and lower sidebands respectively and the last inequality is exactly the quadrature entanglement criterion \cite{DuanPRL00i}.

When tensor polarizability effects can be neglected, $\chi_{p}=\chi_{a}$ and $\xi=0$, the input-output relations reduce to the QND case \cite{HammererRMP09} and entanglement driven by the swap interaction disappears.

The duration of the light mode which contains squeezing can be manipulated by changing the driving power and/or the optical depth of our sample.  The output mode can be also shaped by varying the driving field intensity during the interaction, so that $\swap$ becomes time dependent, which is of particular importance for applications for atomic memories. A closer examination of Eq.~\eqref{Eq:iolight} reveals that the output mode in this case has a mode function $u(t)$ proportional to $ u(t)\propto\swap(t)\exp\left(-\int_0^t dt' \swap(t')\right) $. In particular, one can shape the driving field pulse such that squeezing is produced in a flat top temporal mode.

\section{Experiment}
\begin{figure}
\centering\includegraphics[scale=0.9]{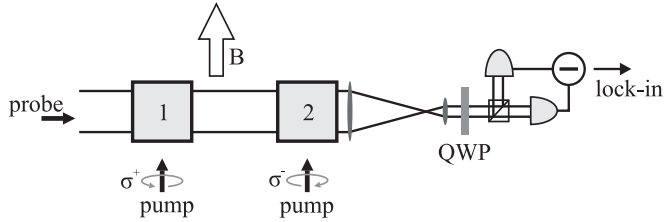}
\caption{Experimental setup. Cells 1 and 2 are placed in magnetic shields in equal magnetic fields $\vec B$ oriented along $x$. The atoms inside are pumped with circularly polarized pump beams of opposite helicity. The probe beam propagates perpendicular to the $\vec B$ field, interacts with the atoms and finally hits a balanced homodyne detector.}
\label{fig:setup}
\end{figure}
The experimental setup is depicted schematically in Fig.~\ref{fig:setup}. We use two 22~mm long cubic paraffin-coated cells containing about $3.6\times10^{11}$ caesium atoms each. Both cells are placed inside magnetic shields in uniform magnetic fields oriented along $x$ axis.
Each measurement cycle begins with optical pumping of the atoms in the two cells into oppositely oriented CSS, as described in \cite{Julsgaard2004,Sherson2006,ShersonRev06}.
  Next a 15~ms long driving pulse blue detuned by 855~MHz from the 6S$_{1/2}$, F=4$\rightarrow$6P$_{1/2}$, F'=5 transition is turned on. Prior to the interaction it is spatially shaped into a circular flat top beam 20~mm in diameter using a telescope beam shaper in order to make the coupling strength uniform across the beam. The driving field polarized along the $y$ axis passes through both cells along the $z$-direction. Finally the beam goes through half- or quarter wave plates, a polarizing beam splitter (PBS) and onto a pair of balanced detectors. Depending on the settings of the wave plates we can either measure the $X_L$ or the $P_L$ quadrature of the light. The signals from the detectors are subtracted, sent to a lock-in amplifier and digitized with an integrating A/D converter at a 12.5~kHz rate.
The number of pumped atoms in the cells is monitored by measuring the Faraday rotation of very weak auxiliary probe beams propagating through the cells along the magnetic field.

\begin{figure}[t]
\centering\includegraphics[scale=0.9]{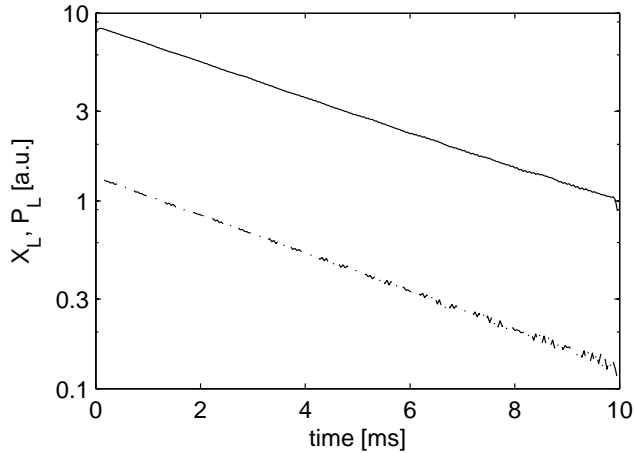}
\caption{The decay of the mean values $\avg{\hat x_L(t)}$ (solid) and $\avg{\hat p_L(t)}$ (dash-dot line) measured by the homodyne detector. Prior to this measurement the atoms are prepared in a CSS state and subsequently displaced in both $\hat X_A$ and $\hat P_A$ by an RF pulse.}
\label{fig:atomdispl}
\end{figure}
As a first test, we check the coupling of the mean value of the atomic spin operators to the light. The initial spin polarized state (CSS) of atoms corresponds to the vacuum state in ($\hat X_A$, $\hat P_A$) phase state. A 150~$\mu$s long RF-pulse applied orthogonally to the dc magnetic field in one of the cells displaces the atomic spins equally in $\hat X_A$ and $\hat P_A$ creating several tens coherent atomic excitations. After that we detect first $\hat X_L$ and then $\hat P_L$ in two series of 200 experimental cycles.
From Eq.~\eqref{Eq:iolight} we can find the expected mean values of the output light operators assuming they have zero mean values at the input:
\begin{align} \label{Eq:meanlight}
\avg{\hat x'_L(t)}&=\frac{\sqrt{2\swap}}{\xi} e^{-\swap t} \avg{\hat P_A(t=0)}, \nonumber \\
\avg{\hat p'_L(t)}&=- \sqrt{2\swap} \xi e^{-\swap t} \avg{\hat X_A(t=0)}.
\end{align}
where the time-dependent light quadrature operators $x'_L(t)+ip'_L(t)=\sqrt2 a''_c(t)$ are written with small letters to distinguish them from operators associated with exponential modes.
The experimental data showing the exponential decay of both mean values as predicted by Eq.\ \eqref{Eq:meanlight} is presented in Fig.~\ref{fig:atomdispl}.
 The ratio $\avg{\hat x_L(0)}/\avg{\hat p_L(0)} $ found from the figure yields $\xi^{-2}=6.3$. 
It agrees very well with the theoretical value at our detuning, which itself is not very sharp due to the Doppler broadening. Notice that for a pure QND interaction, $\avg{\hat p_L(t)}$ would be zero independently of either measurement time or input mean values of the atomic operators.

\begin{figure}[t]
\centering\includegraphics[scale=0.9]{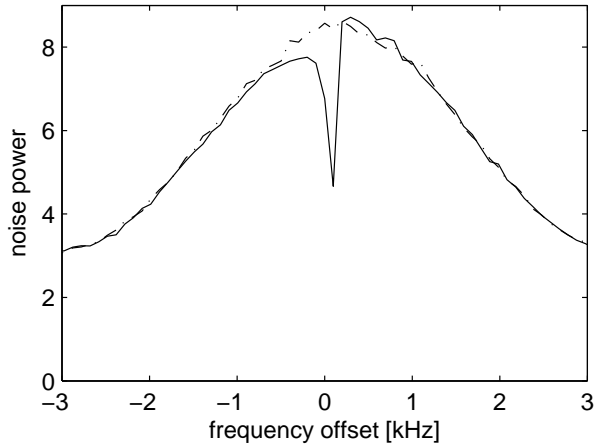}
\caption{Average power spectrum $\avg{\hat p^2_{L,c}(\Omega)+\hat p^2_{L,s}(\Omega)}=\text{Var}(\hat p_{+\Omega}+\hat p_{-\Omega})+\text{Var}(\hat x_{+\Omega}-\hat x_{-\Omega})$ of the homodyne signal $\hat p_L(t)$. Frequencies around $\Omega_L=322$kHz are shown.  The atoms are initialised to the CSS state and then a 15~ms long probe pulse is shined through them. Solid curve was taken with atoms in both cells tuned to the 322~kHz Larmor frequency, while dash-dot curve is the shot noise level reference, taken with atoms detuned far away from the detection bandwidth. The Gaussian shape of the reference spectrum is due to the detection bandwidth, mainly limited by the lock-in amplifier used to demodulate the homodyne signal, while the dip in the middle is the fingerprint of the ultranarrowband squeezing produced in the interaction with the atoms.}
\label{fig:specsqz}
\end{figure}
Next we proceed to the demonstration of the squeezing of the output light. We prepare the atoms in the CSS, send a 15~ms probe pulse and measure the noise in $\hat p_{L,c}(t)$ and $\hat p_{L,s}(t)$, averaging over typically 10000 cycles. Squeezing can be seen in the power spectrum of the signal (i.e. $\avg{\hat p^2_{L,c}(\Omega)+\hat p^2_{L,s}(\Omega)}$) from the lock-in amplifier as shown in Fig.~\ref{fig:specsqz}. When the magnetic field is shifted such that the atomic contribution to the noise lies outside the detection bandwidth, we measure the shot noise level, with the spectral shape corresponding to the gain/sensitivity function of our detection system. When the magnetic fields in both cells are adjusted so that the atoms precess at exactly 322~kHz (the center frequency of the detection range), we see an apparent dip in the noise power at that frequency and around it in the bandwidth of a few hundred hertz.

\section{Analysis of the temporal modes of the squeezing}
The probe light emerging from the cells has a strong vertically polarised component accompanied by a squeezed state in a horizontally polarised component. The latter is typically contained in a temporal exponentially falling mode. Since all the squeezing is contained in a narrow frequency bandwidth readily detectable by our homodyne setup, it is possible to find the excited modes directly from the experimental data. In each experimental cycle we sample the output of the homodyne detector at a rate much higher than necessary to capture the bandwidth in which the squeezing occurs. This way we obtain values of $\hat p$ quadratures of light as a function of time, $\hat p_{L,c}(t)$ and $\hat p_{L,s}(t)$. We focus on the cosine quadrature with the results for the sine quadrature being very similar. The two-time correlation function $\avg{\hat p_{L,c}(t) \hat p_{L,c}(t')}$ yields the amount of noise in any temporal mode or a correlation between pairs of modes that are contained within the detection bandwidth. Therefore it is natural to ask whether any mode basis is favoured in such case. The  answer is provided by the Karhunen-Lo\'eve theorem as detailed in the Appendix. For each correlation matrix one can find a unique basis of modes $u_n(t)$ which have no cross-correlations. Each of them can be in principle measured separately and it will exhibit a variance of $\xi^2_n=\text{Var}(\hat P^{(n)}_{L,c})$ that is also found from the  $\avg{p_{L,c}(t) p_{L,c}(t')}$.

\begin{figure}[t]
\centering\includegraphics[scale=0.9]{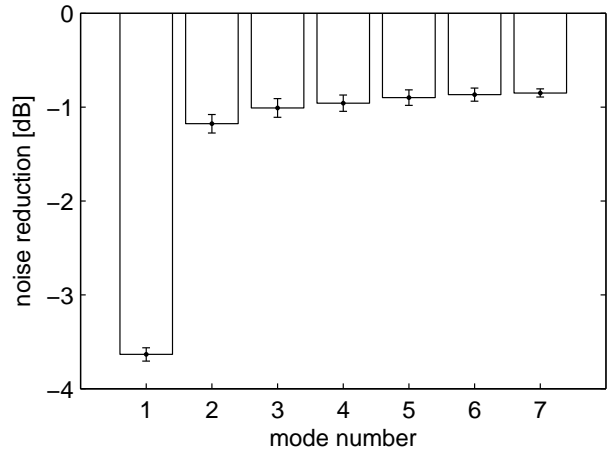}
\caption{Noise reduction  $10\log_{10}\left(2\avg{(\hat P^{(n)}_{L,c})^2+ (\hat P^{(n)}_{L,s})^2}\right)$
in the characteristic output modes.}
\label{fig:eigsqz}
\end{figure}
\begin{figure}[t]
\centering\includegraphics[scale=0.9]{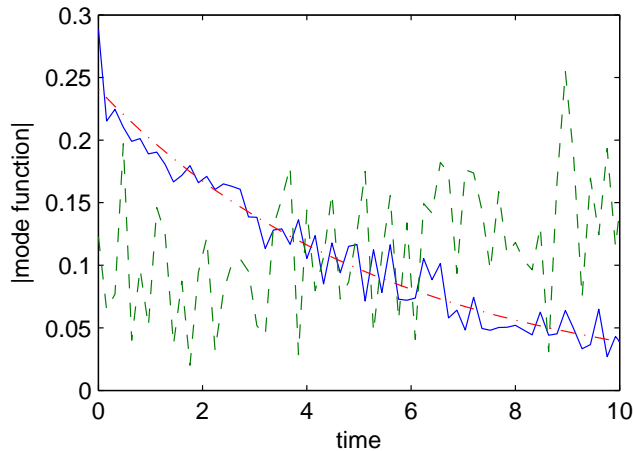}
\caption{Characteristic output mode functions for the most squeezed mode (blue solid line), the next mode (green dashed line) and an exponential fit for the most squeezed mode (red dash-dot line).}
\label{fig:sqzmode}
\end{figure}
In Fig.~\ref{fig:eigsqz} we plot the sum of sine and cosine quadrature variances $2\text{Var}(\hat P^{(n)}_{L,c})+ 2\text{Var}(\hat P^{(n)}_{L,s})
=\text{Var}(\hat P^{(n)}_+ + \hat P^{(n)}_-)/2+\text{Var}(\hat X^{(n)}_+ - \hat X^{(n)}_-)/2$  obtained from the measured correlation matrices. It is expressed as a noise reduction below shot noise. We can see that there is one dominant squeezed mode, in which the noise is reduced by 3.5(1)dB. Superscript $(n)$ denotes $n$-th characteristic mode function $u_n(t)$. Several other modes are squeezed by about 1~dB. This is due to atomic decoherence taking place during the interaction. The equations \eqref{Eq:iotime} describe a perfectly coherent evolution and predict pure single mode squeezing. However, in reality the transverse decoherence time $T_2$ is comparable to the duration of the fundamental mode. Thus, figuratively speaking, while the squeezed light leaves the cells, the state of atoms in the cells is driven back to the initial CSS state. This starts the process of squeezing of the light anew in an incoherent manner. This is confirmed by the shape of the characteristic mode functions obtained from the experiment shown in Fig.~\ref{fig:sqzmode}. The most squeezed mode has an exponentially decaying shape, which agrees with the theoretical model. The decay rate is found to be $(5.5\text{ms})^{-1}$, which is virtually equal to the decay rate of the mean values measured to be $\swap=(5.7\text{ms})^{-1}$. The next mode is rising, supporting the explanation that as the atomic state is brought back to the CSS, an independent squeezing process starts.

\section{Conclusions}
We have developed a model of an off-resonant interaction between light and spin polarized multi-level atomic vapour. We find, that for long interaction times and/or small detunings, the interaction leads to the swap of states between the light and atoms accompanied by two-mode squeezing (entanglement) transformation of the two sidebands of light and the two cells. In particular if the two atomic ensembles are in a CSS prior to the interaction, this state will be mapped with entanglement onto the state of the $\Omega_{L}$ sidebands of the outgoing light. This state of light is emitted in an exponentially decaying temporal mode.
At the same time a portion of incoming light, which comes in an exponentially rising temporal mode is mapped onto the atoms.


We have confirmed experimentally that a room temperature Cs vapour generates a single temporal light mode in the quadrature entangled state with 3.5(1)dB of entanglement. As predicted by theory the temporal mode in which we find entanglement is decaying exponentially. Losses of light and decoherence of the atomic state populate a few other uncorrelated modes weakly squeezed by about 1 dB. We find experimentally that the rate of the leading  decoherence process scales with optical power and the density of atoms approximately in the same way as the rate of the coherent interaction $\swap$. We attribute this decay which currently limits the degree of entanglement to light-induced collisions. Applying antireflection  coating on the cell windows to reduce the losses and reducing atomic decoherence due to collisions and magnetic dephasing should allow to generate even more pure entangled state with a higher degree of squeezing.
We expect that the single mode squeezing and two mode quadrature entanglement can be beneficial in QIP protocols where security is a concern. We also expect that it can be very useful in protocols where discrimination of a single mode at the detection stage is not possible. Note, that the squeezing produced is readily compatible with the atomic memories \cite{Julsgaard2004} and the mode in which it is produced can be shaped by changing the intensity of the local oscillator during the interaction.


This research was funded by EU grants COMPAS, QAP, and HIDEAS. C.M. acknowledges support from the Elite Network of Bavaria, QCCC. Paraffin coating of the cells was skillfully performed by M. Balabas. 
\appendix

\section{Details of the derivation}
It is possible to derive the Hamiltonian from Eq.~\eqref{Eq:HQND+} by evaluating the Clebsh-Gordon coefficients and adiabatically eliminating excited levels. Without further approximations this leads to a Hamiltonian describing the scalar, vectorial and tensorial polarizability of the atoms, with coupling constants of $a_0$, $a_1$ and $a_2$ respectively  \cite{ShersonRev06}:
\begin{align}\label{App:Hint}
\hat H_\text{int}=&\frac{\hbar\gamma}{8\Delta}\frac{\lambda^2}{2\pi} \frac{N_a}{AL}\int_0^L dz \Bigl[
a_0 \hat\phi + a_1 S_z j_z  
+a_2 \left( \phi j^2_z -S_- j_+^2 - S_+ j_-^2\right) \Bigr].
\end{align}
Above $z$ is the direction of the propagation of the light beam, $A$ is the cross section of the beam, $\gamma$ is the natural linewidth while  $\Delta$ is the detuning.
The Hamiltonian \eqref{App:Hint} can be simplified if the atoms are in a state very close to the CSS oriented along $x$ axis. The result is different depending on the $F$ number. For concretness we will assume here $F=4$ with almost all atoms in $m_F=4$ state with respect to the $x$-axis, only some in $m_F=3$ and none in the other states. In this situation we can approximate $j_z^2$, $j_+^2$ and $j_-^2$ by c-numbers and components of $\hat{\bm j}$. This way we arrive at:
\begin{align}
\hat H_{int}=&\frac{\hbar\gamma}{8\Delta}\frac{\lambda^2}{2\pi}\frac{N_a}{AL}
\int_0^L dz \Biggl\{ a_1 S_z j_z - 14 a_2 ( S_y j_y +2 S_x j_x)
+ \hat\phi\left[a_0-a_2\left(16-\frac{7}{2}j_x\right)\right] + 56 a_2 S_x \Biggr\}
\end{align}
The Hamiltonian above may appear complicated, but in fact only two first terms under the integral $a_1 S_z j_z - 14 a_2 S_y j_y$ are nontrivial. The other terms generate classical rotations, in particular: the $S_xj_x$ term in our settings, when both $S_x$ and $j_x$ are macroscopic merely causes a rotation in the $x$--$p$ plane for both light and atoms. The total angle of this rotation is typically of the order of a few miliradians and we neglect it.  $\hat\phi$ is a constant of motion in our system, thus the term $\hat\phi j_x$ only shifts the Larmor frequency --- it represents the Stark shift.  The last term, $S_x$, adds to a rotation of in the $x_c$--$p_c$ plane for light but is still negligible.

Finally we can rewrite nontrivial terms from the above Hamiltonian $a_1 S_z j_z - 14 a_2 S_y j_y$ using bosonic operators, $\hat a$ for light and $\hat b$ for atoms. This is accomplished using the relations:
\begin{align}
S_y&=\frac{\sqrt{\Phi}}{2} (\hat a + \hat a^\dagger) &
S_z&=-\frac{\sqrt{\Phi}}{2i} (\hat a - \hat a^\dagger)
\nonumber\\
j_y&=\sqrt{\frac{2L}{N_a}} (\acute{\hat b} + \acute{\hat b}^\dagger) &
j_z&=-i\sqrt{\frac{2L}{N_a}} (\acute{\hat b} - \acute{\hat b}^\dagger)
 \nonumber\\
\end{align}
where $\Phi$ is the photon flux per unit time and $N_a$ is the total number of atoms and the sign in $S_z$ is a consequence of negative sign of $S_x$ in our settings.
This way we can approximate the interaction Hamiltonian in the form:
\begin{align}
\hat H^\text{appr}_\text{int}=&
\frac{\hbar}{\sqrt{L}}\sqrt{2\swap}\int_0^L dz 
\left[
   \frac{1}{\xi}\frac{\acute{\hat b}-\acute{\hat b}^\dagger}{i\sqrt 2}\frac{\hat a-\hat a^\dagger}{i\sqrt 2}+
   \xi \frac{\acute{\hat b}+\acute{\hat b}^\dagger}{\sqrt 2}\frac{\hat a+\hat a^\dagger}{\sqrt 2}
\right],
\end{align}
where
\begin{align}
\swap&=14 a_1 a_2 \frac{\Phi  N_a}{A^2} \left(\frac{\gamma}{8\Delta}\frac{\lambda^2}{2\pi}\right)^2, &
\xi&=\sqrt\frac{14a_2}{a_1}.
\end{align}
The expressions for the vector $a_{1}$ and $a_2$ tensor polarizabilities for Cs D2 line can be found, for example in \cite{HammererRMP09}. It can be directly verified, that the above Hamiltonian is identical with Eq.~\eqref{Eq:Htot},
with $\chi_a=\sqrt{\swap/2}(1/\xi-\xi)$ and $\chi_p=\sqrt{\swap/2}(1/\xi+\xi)$.

Let us note that the sign of $\xi^2$ can be flipped, that is the $\chi_a$ and $\chi_p$ can be interchanged.
This is accomplished by rotating the polarisation of the driving field by 90$^\circ$, $S_x \rightarrow -S_x$ or by switching from blue to red detuning $\Delta \rightarrow -\Delta$. The input-output relations given in Eq.~\eqref{Eq:io} remain valid for both signs of $\xi^2$. However for a real $\xi$ they describe the entanglement between two sidebands of the light field, whereas for imaginary $\xi$ they entail the entanglement between the light and atoms.


\section{Eigenmode decomposition}
In the experiment we measure directly the covariance matrix of the homodyne signal $p_c(t)$:
\begin{equation}
C(t,t')=\avg{p_c(t) p_c(t')}-\avg{p_c(t)} \avg{p_c(t')}
\end{equation}
where $p_c(t)$ is a properly scaled signal directly form the lock-in amplifier.
After the measurement we can calculate the amount of noise in any temporal mode characterised by the modefunction $u_n(t)$, where $n$ indexes a set of modes we are interested in. A $P$-quadrature operator for the $n$th mode is simply $\hat Q_n=\int u_n(t) p_c(t)$. We can calculate both the variance of any $Q_n$ and correlation between any two of them:
\begin{equation}
\avg{Q_n Q_m}-\avg{Q_n}\avg{Q_m}=\iint dt dt' u_n(t)C(t,t')u_m(t')
\end{equation}
According to the Karhunen-Lo\'eve theorem one can find a set of mutually uncorrelated modes $u_n(t)$ by simply performing a spectral decomposition of measured $C(t,t')$. In this way we obtain eigenvalues $\xi_n$ and eigenfunctions $u_n(t)$:
\begin{equation}
C(t,t')=\sum_n \xi_n u_n(t)u_n(t')
\end{equation}
The eigenvalues $\xi_n$ are equal to the quadrature variances, $\xi_n=\avg{Q_n^2}-\avg{Q_n}^2$, while the eigenfunctions give the quadrature mode functions. The same procedure can be repeated for the sine component of the homodyne signal, $p_s(t)$ and it yields results identical to within the experimental uncertainties.

\end{document}